

%
\catcode`\@=11 

\newcount\yearltd\yearltd=\year\advance\yearltd by -1900

%
%

\def\draftmode{\message{ DRAFTMODE }\def\draftdate{{\rm preliminary draft:
\number\month/\number\day/\number\yearltd\ \ \hourmin}}%
\headline={\hfil\draftdate}\writelabels\baselineskip=20pt plus 2pt minus 2pt
 {\count255=\time\divide\count255 by 60 \xdef\hourmin{\number\count255}
  \multiply\count255 by-60\advance\count255 by\time
  \xdef\hourmin{\hourmin:\ifnum\count255<10 0\fi\the\count255}}}
\def\nolabels{\def\wrlabel##1{}\def\eqlabel##1{}\def\reflabel##1{}}
\def\writelabels{\def\wrlabel##1{\leavevmode\vadjust{\rlap{\smash%
{\line{{\escapechar=` \hfill\rlap{\sevenrm\hskip.03in\string##1}}}}}}}%
\def\eqlabel##1{{\escapechar-1\rlap{\sevenrm\hskip.05in\string##1}}}%
\def\thlabel##1{{\escapechar-1\rlap{\sevenrm\hskip.05in\string##1}}}%
\def\reflabel##1{\noexpand\llap{\noexpand\sevenrm\string\string\string##1}}}
\nolabels
%
\global\newcount\secno \global\secno=0
\global\newcount\meqno \global\meqno=1
\global\newcount\mthno \global\mthno=1
\global\newcount\mexno \global\mexno=1
\global\newcount\mquno \global\mquno=1

\def\newsec#1{\global\advance\secno by1\message{(\the\secno. #1)}
\global\subsecno=0\xdef\secsym{\the\secno.}\global\meqno=1\global\mthno=1
\global\mexno=1\global\mquno=1
\bigbreak\medskip\noindent{\bf\the\secno. #1}\writetoca{{\secsym} {#1}}
\par\nobreak\medskip\nobreak}
\xdef\secsym{}
\global\newcount\subsecno \global\subsecno=0
\def\subsec#1{\global\advance\subsecno by1\message{(\secsym\the\subsecno. #1)}
\bigbreak\noindent{\it\secsym\the\subsecno. #1}\writetoca{\string\quad
{\secsym\the\subsecno.} {#1}}\par\nobreak\medskip\nobreak}
\def\appendix#1#2{\global\meqno=1\global\mthno=1\global\mexno=1%
\global\mquno=1
\global\subsecno=0
\xdef\secsym{\hbox{#1.}}
\bigbreak\bigskip\noindent{\bf Appendix #1. #2}\message{(#1. #2)}
\writetoca{Appendix {#1.} {#2}}\par\nobreak\medskip\nobreak}
%
%
\def\eqnn#1{\xdef #1{(\secsym\the\meqno)}\writedef{#1\leftbracket#1}%
\global\advance\meqno by1\wrlabel#1}
\def\eqna#1{\xdef #1##1{\hbox{$(\secsym\the\meqno##1)$}}
\writedef{#1\numbersign1\leftbracket#1{\numbersign1}}%
\global\advance\meqno by1\wrlabel{#1$\{\}$}}
\def\eqn#1#2{\xdef #1{(\secsym\the\meqno)}\writedef{#1\leftbracket#1}%
\global\advance\meqno by1$$#2\eqno#1\eqlabel#1$$}
%
%
\def\thm#1{\xdef #1{\secsym\the\mthno}\writedef{#1\leftbracket#1}%
\global\advance\mthno by1\wrlabel#1}
\def\que#1{\xdef #1{\secsym\the\mquno}\writedef{#1\leftbracket#1}%
\global\advance\mquno by1\wrlabel#1}
\def\exm#1{\xdef #1{\secsym\the\mexno}\writedef{#1\leftbracket#1}%
\global\advance\mexno by1\wrlabel#1}
%
\newskip\footskip\footskip14pt plus 1pt minus 1pt 
\def\f@@t{\baselineskip\footskip\bgroup\aftergroup\@foot\let\next}
\setbox\strutbox=\hbox{\vrule height9.5pt depth4.5pt width0pt}
\global\newcount\ftno \global\ftno=0
\def\foot{\global\advance\ftno by1\footnote{$^{\the\ftno}$}}
%
\newwrite\ftfile
\def\footend{\def\foot{\global\advance\ftno by1\chardef\wfile=\ftfile
$^{\the\ftno}$\ifnum\ftno=1\immediate\openout\ftfile=foots.tmp\fi%
\immediate\write\ftfile{\noexpand\smallskip%
\noexpand\item{f\the\ftno:\ }\pctsign}\findarg}%
\def\footatend{\vfill\eject\immediate\closeout\ftfile{\parindent=20pt
\centerline{\bf Footnotes}\nobreak\bigskip\input foots.tmp }}}
\def\footatend{}
%
%
\global\newcount\refno \global\refno=1
\newwrite\rfile
\def\ref{\the\refno\nref}
\def\bref{\nref}
\def\nref#1{\xdef#1{\the\refno}\writedef{#1\leftbracket#1}%
\ifnum\refno=1\immediate\openout\rfile=refs.tmp\fi
\global\advance\refno by1\chardef\wfile=\rfile\immediate
\write\rfile{\noexpand\item{[#1]\ }\reflabel{#1\hskip.31in}\pctsign}\findarg}
\def\findarg#1#{\begingroup\obeylines\newlinechar=`\^^M\pass@rg}
{\obeylines\gdef\pass@rg#1{\writ@line\relax #1^^M\hbox{}^^M}%
\gdef\writ@line#1^^M{\expandafter\toks0\expandafter{\striprel@x #1}%
\edef\next{\the\toks0}\ifx\next\em@rk\let\next=\endgroup\else\ifx\next\empty%
\else\immediate\write\wfile{\the\toks0}\fi\let\next=\writ@line\fi\next\relax}}
\def\striprel@x#1{} \def\em@rk{\hbox{}}

\def\addref#1{\immediate\write\rfile{\noexpand\item{}#1}} 
\def\startrefs#1{\immediate\openout\rfile=refs.tmp\refno=#1}
\def\xref{\expandafter\xr@f}\def\xr@f[#1]{#1}
\def\refs#1{[\r@fs #1{\hbox{}}]}
\def\r@fs#1{\edef\next{#1}\ifx\next\em@rk\def\next{}\else
\ifx\next#1\xref #1\else#1\fi\let\next=\r@fs\fi\next}
%

%
\newwrite\ffile\global\newcount\figno \global\figno=1
\def\fig{fig.~\the\figno\nfig}
\def\nfig#1{\xdef#1{fig.~\the\figno}%
\writedef{#1\leftbracket fig.\noexpand~\the\figno}%
\ifnum\figno=1\immediate\openout\ffile=figs.tmp\fi\chardef\wfile=\ffile%
\immediate\write\ffile{\noexpand\medskip\noexpand\item{Fig.\ \the\figno. }
\reflabel{#1\hskip.55in}\pctsign}\global\advance\figno by1\findarg}
\def\xfig{\expandafter\xf@g}\def\xf@g fig.\penalty\@M\ {}
\def\figs#1{figs.~\f@gs #1{\hbox{}}}
\def\f@gs#1{\edef\next{#1}\ifx\next\em@rk\def\next{}\else
\ifx\next#1\xfig #1\else#1\fi\let\next=\f@gs\fi\next}
\newwrite\lfile
{\escapechar-1\xdef\pctsign{\string\%}\xdef\leftbracket{\string\{}
\xdef\rightbracket{\string\}}\xdef\numbersign{\string\#}}

\def\writestop{\def\writestoppt{\immediate\write\lfile{\string\pageno%
\the\pageno\string\startrefs\leftbracket\the\refno\rightbracket%
\string\def\string\secsym\leftbracket\secsym\rightbracket%
\string\secno\the\secno\string\meqno\the\meqno}\immediate\closeout\lfile}}
\def\writestoppt{}\def\writedef#1{}
\def\seclab#1{\xdef #1{\the\secno}\writedef{#1\leftbracket#1}\wrlabel{#1=#1}}
\def\subseclab#1{\xdef #1{\secsym\the\subsecno}%
\writedef{#1\leftbracket#1}\wrlabel{#1=#1}}
\newwrite\tfile \def\writetoca#1{}
\def\leaderfill{\leaders\hbox to 1em{\hss.\hss}\hfill}
\def\writetoc{\immediate\openout\tfile=toc.tmp
   \def\writetoca##1{{\edef\next{\write\tfile{\noindent ##1
   \string\leaderfill {\noexpand\number\pageno} \par}}\next}}}

%
\catcode`\@=12 
%
\ifx\answ\bigans
 
 \font\titlei=cmmi10 scaled\magstep3
\font\titleis=cmmi7 scaled\magstep3 \font\titleiss=cmmi5 scaled\magstep3
\font\titlesy=cmsy10 scaled\magstep3 \font\titlesys=cmsy7 scaled\magstep3
\font\titlesyss=cmsy5 scaled\magstep3 
\else
 
 \font\titlei=cmmi10 scaled\magstep4
\font\titleis=cmmi7 scaled\magstep4 \font\titleiss=cmmi5 scaled\magstep4
\font\titlesy=cmsy10 scaled\magstep4 \font\titlesys=cmsy7 scaled\magstep4
\font\titlesyss=cmsy5 scaled\magstep4 
 
 \font\absi=cmmi10 scaled\magstep1
\font\absis=cmmi7 scaled\magstep1 \font\absiss=cmmi5 scaled\magstep1
\font\abssy=cmsy10 scaled\magstep1 \font\abssys=cmsy7 scaled\magstep1
\font\abssyss=cmsy5 scaled\magstep1 
\skewchar\absi='177 \skewchar\absis='177 \skewchar\absiss='177
\skewchar\abssy='60 \skewchar\abssys='60 \skewchar\abssyss='60
\fi
\skewchar\titlei='177 \skewchar\titleis='177 \skewchar\titleiss='177
\skewchar\titlesy='60 \skewchar\titlesys='60 \skewchar\titlesyss='60
\ifx\answ\bigans\else
 \fi
%

%
%
%

\def\p{\partial}

\def\vev#1{\langle #1 \rangle}

\def\darr#1{\raise1.5ex\hbox{$\leftrightarrow$}\mkern-16.5mu #1}
\def\half{{\textstyle{1\over2}}} 
%
%
\def\al{\alpha}
\def\be{\beta}
\def\ga{\gamma}  
\def\de{\delta}  \def\De{\Delta}

\def\et{\eta}

\def\ka{\kappa}
\def\la{\lambda}

\def\ph{\phi}    

\def\ps{\psi}  
  
%
%

%

%
%

\def\cF{{\cal F}}

\def\cL{{\cal L}}

%
%
%
\def\lefthook{{\vrule height5pt width0.4pt depth0pt}}
\def\righthook{{\vrule height5pt width0.4pt depth0pt}}
\def\leftrighthookfill{$\mathsurround=0pt \mathord\lefthook
     \hrulefill\mathord\righthook$}
\def\underhook#1{\vtop{\ialign{##\crcr$\hfil\displaystyle{#1}\hfil$\crcr
      \noalign{\kern-1pt\nointerlineskip\vskip2pt}
      \leftrighthookfill\crcr}}}
%
%


\def\ie{{\it i.e.\ }}
\def\eg{{\it e.g.\ }}

\def\ZZ{Z\!\!\!Z} 		
\def\NN{I\!\!N}			%
\def\RR{I\!\!R}			
\def\CC{I\!\!\!\!C}		%

\def\bs{\bigskip}



\def\mapright#1{\smash{\mathop{\longrightarrow}\limits^{#1}}}

%


\magnification=1200


\def\fancyhatg{{\rm Hom\,}_{{\cal U}( {\hbox{{\hbox{$\scriptstyle\bf g$}}
\kern-.8em\lower.5ex\hbox{$\scriptstyle\widehat{\phantom{\bf g}}$}}})}}


\def\fff{\cF(p^M,p^L)}

\def\whd{\widehat{d}}

\def\rel{{rel}} \def\abs{{abs}}

\def\whd{\widehat{d}}
\def\NO#1{:\!{#1}\!:}

\def\NP{Nucl. Phys.\ }
\def\PL{Phys. Lett.\ }
\def\CMP{Comm. Math. Phys.\ }

\def\MPL{Mod. Phys. Lett.\ }

\def\LNM{Lect. Notes Math.\ }
\bref\Da{
F. David, {\it Conformal field theories coupled to 2-D gravity
in the conformal gauge}, \MPL {\bf A3} (1988) 1651.}
\bref\DK{
J. Distler and H. Kawai, {\it Conformal field theory and 2-D
quantum gravity}, \NP {\bf B321} (1989) 509.}
\bref\LZb{
B.H. Lian and G.J. Zuckerman, {\it New selection rules and physical
states in 2D gravity; conformal gauge}, \PL {\bf 254B} (1991) 417.}
\bref\BMPa{
P. Bouwknegt, J. McCarthy and K. Pilch, {\it BRST analysis of physical
states for $2D$ gravity coupled to $c<1$ matter}, CERN-TH.6162/91.}
\bref\Fe{
G. Felder, {\it BRST approach to minimal models},
\NP {\bf B317} (1989) 215; erratum, {\it ibid.}  {\bf B324} (1989) 548.}
\bref\BMPb{
P. Bouwknegt, J. McCarthy and K. Pilch, {\it Fock space resolutions of the
Virasoro highest weight modules with $c\leq1$}, CERN-TH.6196/91.}
\bref\LZd{
B.L. Lian and G.J. Zuckerman, {\it $2D$ gravity with $c=1$ matter}, \PL
{\bf 266B} (1991) 21.}
\bref\Poc{
A.M. Polyakov, {\it Selftuning fields and resonant correlations in $2D$
gravity}, \MPL {\bf A6} (1991) 635.}
\bref\GKN{
D.J. Gross, I.R. Klebanov and M.J. Newman, {\it The two point
correlation function of the one-dimensional matrix model},
\NP {\bf B350} (1991) 621.}
\bref\FFu{
B.L. Feigin and D.B. Fuchs,  {\it Representations of the Virasoro algebra},
 in {\it Representations of infinite-dimensional
Lie groups and Lie algebras}, Gordon and Breach, New York (1989). }
\bref\Wi{
E. Witten, {\it Ground ring of two dimensional string theory},
preprint IASSNS-HEP-91/51.}
\bref\MMS{
S. Mukherji, S. Mukhi and A. Sen, {\it Null vectors and extra states in $c=1$
string theory}, TIFR/TH/91-25, May '91.}
\bref\KlPo{
I. Klebanov and A. Polyakov, {\it Interaction of discrete states
in two-dimensional string theory}, preprint PUPT-1281.}
\bref\Gold{
J. Goldstone, unpublished.}
\bref\Kac{V.G. Kac, {\it Contravariant form for infinite dimensional Lie
algebras and superalgebras}, \LNM {\bf 94} (1979) 441.}
\bref\Itb{
K. Itoh, {\it $SL(2,\RR)$ current algebra and spectrum in two
dimensional gravity}, CTP-TAMU-42/91.}
\bref\BK{
M. Bershadsky and I. Klebanov, {\it Genus-one path integral in
two-dimensional quantum gravity},
Phys. Rev. Lett. {\bf 65} (1990) 3088.}
\bref\Br{
R.~Brooks, {\it Residual symmetries in $c=1$ noncritical string theory},
MIT-CTP \#1957.}
\bref\FFK{
G. Felder, J. Fr\"ohlich and G. Keller, {\it Braid matrices and
structure constants for minimal conformal models},
\CMP {\bf 124} (1989) 647.}
\bref\BT{
R. Bott and L.W.\ Tu, {\it Differential forms  in algebraic
topology}, Springer-Verlag, New York, 1982.}
\bref\FK{
J.~Figueroa-O'Farrill and T. Kimura, {\it The BRST cohomology of the NSR
string: vanishing and ``no ghost'' theorem}, \CMP {\bf 124} (1989) 105.}
\bref\LZc{
B.H. Lian and G.J. Zuckerman, {\it BRST cohomology of the super-Virasoro
algebras}, \CMP  {\bf 125} (1989) 301.}
\bref\BKT{
M. Bershadsky, V. Knizhnik and M. Teitelman, {\it Superconformal symmetry
in two dimensions},
\PL {\bf 151B} (1985) 31.}
\bref\IMNU{
M. Ito, T. Morozumi, S. Nojiri and S. Uehara, {\it Covariant quantization
of Neveu-Schwarz-Ramond model},
Prog. Theor. Phys. {\bf 75} (1986) 934.}
\bref\Eich{
H. Eichenherr, {\it Minimal operator algebras in superconformal quantum
field theory},
\PL {\bf 151B} (1985) 26.}
\bref\FQS{
D. Friedan, Z. Qiu and S. Shenker, {\it Supercon\-for\-mal in\-vari\-ance in
two-di\-men\-sions and the tricritical Ising model},
\PL {\bf 151B} (1985) 37.}
\bref\IO{
K. Itoh and N. Ohta, {\it BRST cohomology and physical states in $2D$
supergravity coupled to $\hat c\leq1$ matter}, FERMILAB-PUB-91/228-T.}

\def\tilpage{
\nopagenumbers\pageno=0
\line{\hfill{CERN-TH.6279/91}}
\bigskip
\centerline{{\bf BRST ANALYSIS OF PHYSICAL STATES FOR $2D$ (SUPER) GRAVITY }}
\centerline{{\bf COUPLED TO (SUPER) CONFORMAL MATTER}\footnote{$^*$}{
To appear in the proceedings of the 1991 Carg\`ese Summer School
on ``New Symmetry Principles in Quantum Field Theory,'' Carg\`ese, July
15-27, 1991.  }}
\bigskip\bigskip
\centerline{Peter Bouwknegt}
\smallskip
\centerline{{\it CERN - Theory Division}}
\centerline{{\it CH-1211 Geneva 23}}
\centerline{{\it Switzerland}}
\bigskip
\centerline{Jim McCarthy}
\smallskip
\centerline{{\it Department of Physics}}
\centerline{{\it Brandeis University}}
\centerline{{\it Waltham, MA 02254}}
\bigskip
\centerline{{ Krzysztof Pilch}}
\smallskip
\centerline{{\it Department of Physics}}
\centerline{{\it University of Southern California}}
\centerline{{\it Los Angeles, CA 90089-0484}}
\bigskip
\bigskip
\centerline{{\bf Abstract}}\smallskip
We summarize some recent results on the BRST analysis of physical
states of $2D$ gravity coupled to $c\leq 1$ conformal matter and
the supersymmetric generalization.

\bs\bs\bs

\vfil
\line{ CERN-TH.6279/91   \hfill }
\line{ BRX TH-326 \hfill}
\line{ USC-91/34 \hfill October 1991}

\eject}
\tilpage
%
%
%
\hsize=6truein \vsize=8.5truein
\voffset=.375truein  \hoffset=.35truein

\font\ninerm=cmr9
\font\ninebf=cmbx9
\footline={\hss\tenrm\folio\hss}
\centerline{{\ninebf BRST ANALYSIS
OF PHYSICAL STATES FOR $2D$ (SUPER) GRAVITY}}
\centerline{{\ninebf COUPLED TO (SUPER) CONFORMAL MATTER}}
\vskip.8truecm
\centerline{{\ninerm PETER BOUWKNEGT}}\smallskip
\centerline{{\it CERN-TH, CH-1211 Geneva 23, Switzerland}}
\vskip.5truecm
\centerline{{\ninerm JIM McCARTHY}}\smallskip
\centerline{{\it Department of Physics, Brandeis University }}
\centerline{{\it Waltham, MA 02254, USA}}
\vskip.5truecm
\centerline{{ \ninerm KRZYSZTOF PILCH}}\smallskip
\centerline{{\it Department of Physics, University of Southern California}}
\centerline{{\it Los Angeles, CA 90089-0484, USA}}
\vskip1.3truecm
\vbox{\hbox{\centerline{{\ninerm ABSTRACT}}}
{\smallskip\leftskip 3pc \rightskip 3pc \noindent \ninerm \baselineskip=10pt
We summarize some recent results on the BRST analysis of physical
states of $2D$ gravity coupled to $c\leq 1$ conformal matter and
the supersymmetric generalization.
\smallskip}}

\def\acknow{
\newsec{Acknowledgements}

P.B.\ would like to thank the organizers for an enjoyable and
stimulating conference, and V. Dotsenko, S. Mukhi, J. Sidenius and E. Verlinde
for useful discussions.
The work of J.M.\ was supported by the NSF
Grant \#PHY-88-04561 and the work of K.P.\ was supported in part by
funds provided by the DOE Contract \#DE-FG03-84ER-40168 and by the USC
Faculty Research and Innovation Fund.

}
\baselineskip=1.2\baselineskip

\newsec{Introduction}

If the continuum theory of $2D$ gravity can be developed to the level
achieved by  matrix models, the combined approaches are expected to
provide great insights into generally covariant quantum systems -- one
of particular interest being the critical ``$2D$ string.'' A further
benefit would be progress in $2D$ supergravity models, where matrix
models seem difficult to apply.
 The continuum theory in the conformal gauge [\Da,\DK] has already had
some success, in particular the  computation of the spectrum of
physical states employing a BRST analysis [\LZb], as we will review
here along the lines of [\BMPa].   One of the new results presented in
this paper is the extension of these calculations to $N=1$ supergravity
models.  Another is the result for representations on the boundary of
the Kac table in a given Virasoro minimal model coupled to gravity.
There is a large literature discussing BRST cohomology and/or $2D$
gravity which has greatly influenced the material presented here.  Due
to space limitation we must refer the reader to [\BMPa] for a more
complete account of these references.

Mathematically, the problem is that of computing the cohomology of the
BRST operator \eqn\KPva{ d= \oint {dz\over 2\pi i}\,\NO{(T^M(z)+ T^L(z)
+\half T^G(z) )c(z)}\,, } on the tensor product
$\cL^M\otimes\cF^L\otimes\cF^G$ of modules of the Virasoro algebra (see
\eg [\LZb]). Here, representing the matter sector, $\cL^M $ is an
irreducible highest weight module $\cL(\De^M,c^M)$ for $c^M<1$, or a
free scalar field Fock space $\cF(p^M)$ for the $c^M=1$ string. The
Fock space $\cF^L\equiv\cF(p^L,Q^L)$ of a free scalar field with a
background charge $Q^L$ (\ie a Feigin-Fuchs module) represents the
Liouville mode, and $\cF^G$ is the Fock space of the spin $(2,-1)$
$bc$-ghosts.  Finally, $T^M(z)+T^L(z)$, is the stress energy tensor of
the matter and the Liouville fields, while $T^G(z)$ is that of the
ghosts. We will refer to the cohomology of $d$ as the BRST cohomology
of $\cL^M\otimes \cF^L$.

Any irreducible highest weight module  of the Virasoro algebra admits a
resolution in terms of Feigin-Fuchs modules [\Fe,\BMPb]. Thus we may
first compute the BRST cohomology of a product of two Feigin-Fuchs
modules and then use a suitable resolution to project from the free
field Fock space onto the irreducible representation in the matter
sector.  For the $c^M = 1$ string the second step is of course
unnecessary.

It is particularly interesting, as discussed above, to study this
problem in the supersymmetric category.  One then considers the BRST
cohomology of  a super-Virasoro matter  module coupled to a
super-Liouville system.

This paper is organized as follows. In  Section 2 we briefly summarize
the results of [\LZd,\BMPa] on the BRST cohomology of a product of two
Feigin-Fuchs modules, and discuss explicit representatives of this
cohomology in the case $c^M=1$.  Then, in Section 3, we use recently
constructed resolutions to  determine the cohomology for all matter
modules with $c^M\leq 1$.  Finally, in Section 4  we give several  new
results for $2D$ supergravity coupled to superconformal matter.

\newsec{The cohomology of two Feigin-Fuchs modules }

The Feigin-Fuchs module $\cF(p,Q)$ is the Fock space of a scalar field,
$\phi(z)$, with a background charge $Q$ and momentum $p$.  We take (see
[\BMPa]) $\vev{ \ph(z)\ph(w) }  = - \ln(z-w)\,,$ $i\p\ph(z) =
\sum_{n\in\ZZ} \al_nz^{-n-1}$ and $T(z) = -\half \NO{ \p\ph(z)
\p\ph(z)} + iQ \p^2\ph(z)$.  The Fock space $\cF(p,Q)$ has vacuum
$|p\rangle$, $\al_0|p\rangle=p|p\rangle$.  As a Virasoro module,
$\cF(p,Q)$ has central charge $c= 1-12Q^2$ and conformal dimension
$\De(p)=\half p(p-2Q)$. We will distinguish between the Liouville and
matter fields by writing superscripts $L$ and $M$ respectively.

We will study the cohomology of the BRST operator \KPva\ on
$\cF(p^M,p^L) \equiv \cF(p^M,Q^M) \otimes \cF(p^L,Q^L)\otimes
\cF^G$.  The ghost number (gh) is normalized such that the physical
vacuum $|p^M,p^L\rangle\equiv|p^M\rangle\otimes|p^L\rangle\otimes
|0\rangle_G$ has ${\rm (gh)}=0$, where the ghost vacuum $|0\rangle_G$
is annihilated by the modes $ c_n,b_n$, $n\geq 1$ and by $b_0$. The
nilpotency of $d$ requires that the total central charge is zero, and
thus $(Q^M)^2+(Q^L)^2=-2$.  On defining $\al_\pm=(Q^M\pm
iQ^L)/{\sqrt2}$ we then have $\al_+\al_-=-1$.  Clearly, at this point
the free parameters in the problem are $\al_+$ (or, equivalently,
$\al_-$) determined by the background charges, and the momenta $p^M$
and $p^L$.  Let us parametrize the latter in terms of $r,s\in\CC$ as
follows \eqn\PBl{p^M - Q^M = \sqrt\half (r\al_+ + s\al_-)\,,\qquad i(
p^L-Q^L)  =  \sqrt\half(r\al_+ - s\al_-) \,. }

The BRST operator can be expanded into the ghost zero modes, $d =
c_0L_0 - b_0M + \widehat{d}$,  where $L_0=\{d,b_0\}$ is the energy
operator.  By a standard argument the cohomology of $d$ must be
contained in the  zero eigen\-space of $L_0$.  Moreover, the subspace
of $\cF(p^M,p^L)$ annihilated by $L_0$ {\it and} $b_0$ is invariant
under $d$, which reduces there to $\widehat d$.  The relative
cohomology of $\cF(p^M,p^L)$ is defined as the cohomology of $d$ on
$\cF(p^M,p^L)\cap {\rm Ker}\,L_0\cap {\rm Ker}\,b_0$, as opposed to the
absolute cohomology which refers to the entire space.
We will denote them by $H_{rel}^{(*)}(\cdot,d)$ and
$H_\abs^{(*)}(\cdot,d)$, respectively.

In [\BMPa] (see also [\LZd]) we proved the following theorem which
summarizes all possible cases with nontrivial relative
cohomology.\thm\LZTa

\proclaim Theorem \LZTa.  We can distinguish three different cases
listed below in which the relative cohomology $H^{(*)}_\rel(\fff,d)$ is
nontrivial.  For given $Q^M$ and $Q^L$, or, equivalently, $\al_+$ and
$\al_-$ satisfying $\al_+\al_-=-1$, they depend on discrete values of
the momenta $p^M$ and $p^L$  parametrized by $r$ and $s$ as in \PBl.
\smallskip \item{i)} If either $r=0$ or $s=0$ then
$$H^{(n)}_\rel(\fff,d)  = \cases {\CC & for $ n=0\,,$\cr
			       0 & otherwise $\,.$\cr}$$
\item{ii)} If  $r,s\in\ZZ_+$  then $$H^{(n)}_\rel(\fff,d) = \cases {\CC
& for $ n=0,1\,,$\cr        0 & otherwise $\,.$\cr}$$ \item{iii)} If
$r,s\in\ZZ_-$ then $$H^{(n)}_\rel(\fff,d) = \cases {\CC & for $
n=0,-1\,,$\cr    0 & otherwise $\,.$ \cr}$$

In case {\sl i)} the cohomology state is  at level zero, \ie it is just
the vacuum, and the conformal dimensions of two Fock spaces satisfy
$\De(p^L)=1-\De(p^M)$ [\Da,\DK].
The states corresponding to cases {\sl ii)} and
{\sl iii)} are called  discrete [\Poc,\GKN], and their level is equal to
$rs$.  Finally, we observe that in these two cases both $\cF^M$ and
$\cF^L$ are reducible Virasoro modules [\FFu,\LZd].

The absolute cohomology is given by
\eqn\KPvb{H^{(*)}_\abs(\fff,d)\simeq
H_\rel^{(*)}(\fff,d)\oplus c_0 H_\rel^{(*-1)}(\fff,d)\,.} More
explicitly, each representative  $\psi$ of the relative cohomology
gives rise to two states $\psi$ and $c_0\psi-\chi$ in the absolute
cohomology, with $\chi$ satisfying $M\psi=\widehat d \chi$.

\def\rst{{1\over \sqrt2}(r-s)}

\def\rksk{{1\over\sqrt2}(r_k-s_k)}

To reveal the physical consequences of the above cohomology it is
important to construct explicit representatives, which is a nontrivial
problem for cases {\sl ii)} and {\sl iii)}.  A particularly interesting
case, which we consider here, is $c^M=1$ (\ie $\al_+=-\al_-=1$), where
the operator cohomology with $(r,s)=(-1,-2)$ and $(-2,-1)$ in
$H^{(-1)}_\rel(\fff,d)$ generate a characteristic ring for the theory
[\Wi].  Further, the states in $H^{(0)}_\rel(\fff,d)$ with $r,s \in
\ZZ_-$ are associated to symmetry currents which preserve this ring.

The matter Fock space for $c^M=1$ decomposes into a direct sum of
irreducible Virasoro modules, $\cF(\rst)=\bigoplus_{\ell=0}^\infty
\cL({1\over 4}(|r-s|+2\ell)^2,1)$.  Moreover, the submodules obtained
by restricting this sum to $\ell\geq\l_0\geq 0$ are isomorphic with
$\cF(\rksk)$, where $k= \max(0,s-r)+\ell_0$, $r_k = r+k$, $s_k = s-k$.
Introduce a vertex operator $V^M=(1/2\pi i) \oint dz
\NO{\exp(-i\sqrt2\ph^M(z))}$. This isomorphism is then realized by the
embedding $(V^M)^k:\cF(\rksk)\rightarrow \cF(\rst)$.  It is also clear,
as one can verify using the direct sum decomposition of both matter
Fock spaces, that the extension of $(V^M)^k$ to an embedding of
$\cF(\rksk,p^L)$ into $\cF(\rst,p^L)$ will map a nontrivial cohomology
class in $\cF(\rksk,p^L)$ into a nontrivial cohomology  class in
$\cF(\rst,p^L)$.

This observation can be put to use as follows. To obtain ${\rm (gh)}=0$
representatives for case {\sl ii)} say, we may take $k=s$ so that
$s_k=0$. By Theorem \LZTa\ {\sl i)} the cohomology of $\cF(\rksk,p^L)$
is given by the vacuum state $|\rksk,p^L\rangle$, which is then mapped
into a singular vector in $\cF(\rst,p^L)$.   For ${\rm (gh)}=0$ case
{\sl iii)} we take $k=-r$ so that $r_k = 0$ and proceed in the same
way.  Finally, the construction of the other ghost number
representatives  for cases {\sl ii)} and {\sl iii)}
 is similarly reduced to finding them for $s_k=1$, $r_k\in\ZZ_+$, and
$r_k=-1$, $s_k\in\ZZ_-$, respectively.  This is quite easy to
accomplish.\thm\CRTrepr

Before we state the result,  let us
recall that the elementary Schur polynomials $S_k(x)$, $x=(x_1,x_2,
\ldots\,)$, are defined through their generating function \eqn\schur{
\sum_{k\geq 0}S_k(x)z^k=\exp\Big( \sum_{k\geq 1}x_k z^k\Big)\,,} and
for convenience we put $S_k(x)=0$ for $k<0$.  To any partition (Young
tableau) $\la=\{\la_1\geq \la_2\geq\ldots\}$ we then associate a Schur
polynomial \eqn\schurgen{ S_{\la_1,\la_2,\ldots}(x) =\det\left(
S_{\la_i+j-i}(x)\right)_{i,j} \,.}

\proclaim Theorem \CRTrepr. The following are  representatives
of the discrete states for $c^M=1$\smallskip
\item{ii)} For $r,s\in\ZZ_+$:
$$\eqalign{
\psi_{r,s}^{(0)}&=(V^M)^s|\sqrt\half(r+s),p^L\rangle
= (-1)^{{1\over2}s(s-1)}s!\,
S_{\underbrace{\scriptstyle{r,\ldots,r}}_{{\hbox{$\scriptstyle
 s$}}}}(x) |p^M,p^L\rangle \,,\cr
\psi_{r,s}^{(1)}&=(V^M)^{s-1}\sum_{q=1}^{r+s-1} S_{r+s-1-q}
(-\sqrt2\al_{-j}^M/j)c_{-q}
|\sqrt\half(r+s-2),p^L\rangle\,.\cr}$$
\item{iii)} For $r,s\in\ZZ_-$:
$$\eqalign{
\psi_{r,s}^{(0)}&=(V^M)^{-r}|-\sqrt\half(r+s),p^L\rangle
=  (-1)^{{1\over2}r(r-1)} (-r)!\,
S_{\underbrace{\scriptstyle{-s,\ldots,-s}}_{{\hbox{$\scriptstyle
 -r$}}}}(x) |p^M,p^L\rangle \,,\cr
\psi_{r,s}^{(-1)}&=(V^M)^{-r-1}\sum_{q=1}^{-(r+s+1)}
 S_{-(r+s+1)-q}(-\al^-_{-j}/j)b_{-q}
|-\sqrt\half(r+s+2),p^L\rangle\,,\cr}$$
where $p^M=\sqrt\half(r-s)\,,
ip^L = \sqrt\half(r+s+2)$, $\al^\pm_n=\sqrt\half
(\al^M_n\pm i\al^L_n)$, and $S_{\la_1,\la_2,\ldots}(x)$ are the
Schur polynomials with argument $x_j = -\sqrt2 \al_{-j}^M/j$.

For ghost number zero these explicit formulae for the physical states,
or the simplest examples, were known to many people (see \eg
[\Poc,\MMS,\KlPo]) and can be traced back to [\Gold,\Kac].  For ${\rm
(gh)}=\pm1$ with low values of $r$ and $s$ examples are given in [\Wi].
In case {\sl ii)} one can also  write down  explicit representatives in
terms of Schur polynomials of the light-cone oscillators $\al^+_n$ only
[\BMPa], but we do not know of a comparably succinct representation in
case {\sl iii)}.

Representatives of the cohomology $H^{(n)}_\rel(\fff,d)$ for $c^M<1$ are
easily obtained from the $c^M=1$ representatives
by making use of the $SO(2,\CC)$ symmetry of [\Itb,\LZd].

It is interesting to observe that the pairing of discrete states for
$c^M=1$, which causes their cancellation in the character [\BK], can be
understood from the existence of a ``supersymmetry-like'' operator
$[d,i\ph^M(z)] = c(z) i\p\ph^M(z)$
(discussed in [\Br]). Clearly, this operator maps between
the (absolute) cohomolgies of $d$, and it can be shown that its zero mode
indeed acts nontrivially.

\newsec{BRST cohomology of highest weight irreducible modules}

The projection onto  $\cL(\De^M,c^M)$ in the matter sector can be
implemented by a suitable resolution. Using the results of Feigin and
Fuchs on the submodule structure of the Fock space modules [\FFu], one
can show that for any irreducible highest weight module  $\cL(\De,c)$
there exists a complex of Feigin-Fuchs modules $(\cF^{(n)},\de^{(n)})$
with cohomology $H^{(n)}(\cF,\de)\simeq \de^{n,0}\cL(\De,c)$. Depending
on the type of the irreducible module such a resolution is given by
either a finite or an infinite (double sided) complex.  Spaces
$\cF^{(n)}$ in the complex differ by their momentum, $p^{(n)}$,  but
have the same background charge $Q$.  The differentials $\de^{(n)}$
commute with the action of the Virasoro algebra, and can be constructed
explicitly in terms of screening currents. For an explicit construction
of resolutions for various types of modules with $c\leq 1$ (labeled as
in the Feigin-Fuchs classification [\FFu]) we refer the reader to:
[\Fe] for type $III_-$ in the interior of the Kac table; [\FFK] for
type $II$; [\BMPb] for $III^0_-$ (the boundary of the Kac table),
$III_-$ in the exterior of the Kac table,  and $III^{00}_-$.

Let ${\cal E}(\De,c)= \{\De(p^{(n)})\}$ be the set of conformal
dimensions of Fock spaces labelled by the degree $n$ in the
resolution.  For modules of type $I$, $II$ and $III_-$ it coincides
with $E(\De,c)$, the set of conformal weights of singular vectors
(including the highest weight vector) in the Verma module ${\cal
V}(\De,c)$ (see [\FFu] for explicit formulae).  For modules of type
$III_-^0$ and $III^{00}_-$ it consists of two highest elements of
$E(\De,c)$.  Denote $\widetilde{\cal E}(\De,c)\equiv 1-{\cal
E}(\De,c)$, and for $\De=1-\De(p^{(n)})$ define $d(\De)=|n|$. One
should note that $d(\cdot)$ does not depend on the choice of the Fock
space resolution.

The BRST cohomology of $\cL(\De^M,p^L)\equiv\cL(\De^M,c^M)\otimes
\cF(p^L,Q^L)$ can be computed by considering a double complex
$(\cF^{(*)}\otimes \cF(p^L,Q^L)\otimes \cF^{G},\de,d)$, with two
commuting differentials $\de$ and $d$, and  the corresponding grading
by the degree in the resolution and the ghost number.  The crucial
observation of [\BMPa] is firstly that, since $\cL^*(\De^M,c^M)\simeq
\cL(\De^M,c^M)$, one has a choice between a given resolution or its
dual.  Then, by explicit examination one always finds that, given
$p^L$, there is a choice of resolution such that, when $p^M$ ranges over
the momenta of all Fock spaces in that complex, \PBl\ has no solution
for integral $r$ and $s$ with $rs > 0$.  Then it follows that the
BRST cohomology of  $\cF^{(n)}\otimes \cF(p^L,Q^L)$ is nontrivial for
at at most one degree $n$, say $n=n_0$, where it is given by  case {\sl
i)} of Theorem \LZTa.\foot{By the remark after Theorem \LZTa\ this is
clearly true when $\cF({p^M,Q^M})\simeq \cL(\De^M,c^M)$ is
irreducible.} A standard argument for double complexes [\BT]  then
gives
\eqn\CRcoh{
\eqalign{
H^{(n)}_\rel(\cL(\De^M,p^L),d)&=
H^{(n)}_\rel(H^{(0)}(\cF\otimes\cF(p^L,Q^L)\otimes \cF^G,\de),d)\cr
&=H^{(n)}(H^{(0)}_\rel(\cF\otimes\cF(p^L,Q^L)
\otimes\cF^G,d),\de)\simeq \de^{n,n_0}\,\CC\,.\cr}}

We can summarize the complete result as follows.\thm\CRTirrr

\proclaim Theorem \CRTirrr.   For any  highest weight irreducible
module, the relative BRST cohomology of $\cL(\De^M,c^M)\otimes
\cF(p^L,Q^L)$ is nontrivial if and only if $\De(p^L)\in\widetilde{\cal
E}({\De^M,c^M})$, in which case $H_\rel^{(n)}(\cL(\De^M,c^M)\otimes
\cF(p^L,Q^L))=\de_{n,\eta(p^L)d(\De(p^L))}\,\CC$, where $\eta(p^L)=
{\rm sign}\,i(p^L-Q^L)$.

For the BPZ minimal models (\ie modules in the interior of the Kac
table) this theorem has first been proven by a somewhat different
method in [\LZb], and then rederived as above in [\BMPa].

Let us illustrate our method in the simple example of a representation
corresponding to the boundary of the Kac table. In this case
$\al_+=\sqrt{p'/p}$, $\al_-=-\sqrt{p/p'}$, where $p'\geq p\geq 1$ are
relatively prime integers. The central charge is
$c^M(p,p')=1-6(p-p')^2/pp'$.  Let $p^M(m,m')$ denotes a matter momentum
as in \PBl\ with $r=-m$ and $s=-m'$, $\cF^M(m,m')$ the corresponding
Fock space, and $\De^M(m,m')$ the conformal dimension. We have the
following two (dual to each other) resolutions of the highest weight
module $\cL(\De^M(m,0),c^M(p,p'))$,  $1\leq m\leq p-1$ [\BMPb]
\eqn\CRbrr{
\matrix{0 & \mapright{\quad\de^{(-2)}} & \cF^M(m-2p,0)
           & \mapright{\quad\de^{(-1)}} & \cF^M(m,0)
            & \mapright{\de^{(0)}} & 0\cr}}
\eqn\CRbrp{
\matrix{0 & \mapright{\quad\de^{(-1)}} & \cF^M(-m,0)
           & \mapright{\de^{(0)}} & \cF^M(-m+2p,0)
            & \mapright{\de^{(1)}} & 0\cr}}
Given $p^L$, integers  $r$ and $s$ in \PBl\ are only determined up to
$r\rightarrow r+tp$ and $s\rightarrow s-tp'$, $t\in\ZZ$.  However,
using $p^M(m+p,m'+p')=p^M(m,m')$, one checks by inspection that -- even
with this ambiguity -- in one of the resolutions \CRbrr\ or \CRbrp\ the
$p^M$ obtained from \PBl\ for $rs > 0$ never appear, just as we have
claimed above.  Finally, one should note that whereas the corresponding
Verma module has infinitely many singular vectors, by Theorem \CRTirrr\
the discrete states arise in only three ghost numbers.  Strikingly
different than for the modules in the interior of the Kac table!

\newsec{The super extension}

\def\FFs{\cF^s(p^M,p^L)}
The results of the previous sections can be generalized to $2D$
supergravity coupled to $\hat c\leq1$ superconformal matter (for the
critical superstring see [\FK,\LZc]).  In this case we deal with a Fock
space $\cF^s(p,Q)$ of a scalar field $\ph(z)$ and a fermionic field
$\ps(z)$, labelled by the scalar zero mode $p$ and background charge
$Q$ as in Section~2. The generators of the $N=1$ superconformal algebra
are given by [\BKT]
\eqn\PBEa{
T(z)  = -\half :\p\ph\p\ph: + iQ\p^2\ph - \half
\ps\p\ps\,, \qquad G(z)  = i\p\ph\ps + 2Q\p\ps\,,
}
and the central
charge equals $\hat c= {2\over3}c=1-8Q^2$.  Physical states correspond
to cohomology classes of the BRST operator

\eqn\PBEb{
d = \oint {dz\over 2\pi i} \NO{\left( (T^M +T^L +\half T^G)(z)c(z) +
(G^M + G^L + \half G^G)(z)\ga(z) \right)}\,.
}
In this case, nilpotency requires $\hat{c}^M+\hat{c}^L=10$, \ie
$(Q^M)^2 + (Q^L)^2 =-1$.  First we give the result for the relative
cohomology of $d$ on the Fock space $\cF^s(p^M,p^L) \equiv
\cF^{s}(p^M,Q^M)\otimes\cF^{s}(p^L,Q^L)\otimes
\cF^{sG}$ where $\cF^{sG}$ is
the Fock space of the $bc$-ghosts and their superpartners $\be\ga$.  As
in \PBl\ we parametrize the momenta $(p^M,p^L)$ in terms of $\al_\pm =
(Q^M \pm iQ^L)/\sqrt2$ and $r,s\in\CC$.  Note, however, that now
$\al_+\al_-=-\half$.

In the following, relative cohomology means the cohomology relative to
$\{L_0,b_0\}$ in the Neveu-Schwarz sector ($\ka=\half$), and the
cohomology relative to $\{ L_0, G_0, b_0, \be_0\}$ in
the Ramond sector ($\ka=0$).
\thm\PBETa
\proclaim Theorem \PBETa. Let $(p^M,p^L)$ be parametrized by $(r,s)$ as in
\PBl.
The relative cohomology of $d$ on the Fock space
$\FFs$
is nontrivial only in the following three cases
\item{(i)} If either $r=0$ or $s=0$ (\ie $\De(p^M)+\De(p^L) = \half$), then
$$
H^{(n)}_{rel}(\FFs,d) = \cases{ \CC & if $n=0\,,$ \cr 0 & otherwise$\,.$\cr}
$$
\item{(ii)} If $r,s\in\ZZ_+\,, r-s\in2\ZZ+(1-2\ka)$, then
$$
H^{(n)}_{rel}(\FFs,d) = \cases{ \CC & if $n=0,1\,,$ \cr 0 & otherwise
$\,.$\cr}
$$
\item{(iii)} If $r,s\in\ZZ_-\,, r-s\in2\ZZ+(1-2\ka)$, then
$$
H^{(n)}_{rel}(\FFs,d) = \cases{ \CC & if $n=0,-1\,,$ \cr 0 & otherwise
$\,.$\cr}
$$
\par

Note that ``discrete states'' occur only if both $\cF^{sM}$ and
$\cF^{sL}$ are reducible. They occur at the same level, namely
${rs\over2}$, as where the null vector in these modules occurs.

As in the bosonic case [\BMPa]
the proof can be given by examining the spectral
sequence associated to the filtered complex obtained by assigning the
degree equal $+1$ to the oscillators $\al_n^+$, $c_n$, $\ps_r^+$,
$\ga_r$ and $-1$ to $\al_n^-$, $b_n$, $\ps_r^-$, $\be_r$,
respectively.  In the NS sector, contrary to the bosonic case, the
spectral sequence does not always collapse after the first term (\ie
the cohomology of the lowest degree differential $\whd_0$ is
not always concentrated in a single degree), and one has to calculate
also the second term, which then yields the final result.  In the
R sector there is an additional complication due to the fact that
$G_0$ does not act reducibly on $\FFs$.  This is resolved by
introducing ``rotated" oscillators -- exactly as in [\IMNU] for the
critical string but without the exponential rescaling -- which
effectively diagonalize $G_0$ in a subspace of $\FFs$.  One can show
that this subspace contains all states annihilated by $G_0$, and in
terms of these new oscillators the cohomology computation is parallel
to that in the NS sector.

Explicit representatives of the cohomology can be given in terms of
``super Schur polynomials'', and the physical states are paired by the
action of the zero mode of the operator $[d,i\ph^M(z)] = c(z) i\p\ph^M(z)+
\ga(z)\ps^M(z)$.
Details will appear elsewhere.

Finally, $\hat c<1$ super minimal models are parametrized by
[\Eich,\BKT,\FQS]
\eqn\PBEc{
\hat{c} (p,p') = 1 - 2 { (p-p')^2\over pp'} \,,\quad
\De(m,m')  = { (mp'-m'p)^2 - (p-p')^2\over 8pp'} +
\textstyle{{1\over16}}(1-2\ka), }
where $p,p'\in2\NN-1\,,{\rm gcd}(p,p')=1$ or $p,p'\in2\NN\,, \half(p-p')\in
2\NN-1\,,{\rm gcd}({p\over2},{p'\over2})=1$, and $1\leq m\leq p-1\,,
1\leq m'\leq p'-1\,, m-m'\in2\ZZ  + (1-2\ka)$.

Let $\al_+ = \sqrt{p'/(2p)}$ and $\al_-= - \sqrt{p/(2p')}$.
Define
\eqn\PBEd{\eqalign{
p^{(n)}_\pm(\ell,\ell') & = \cases{ \pm\sqrt\half \left(
 (\ell\pm 2np)\al_++\ell'\al_-\right) & for $n$ even$\,,$ \cr
\pm\sqrt\half \left( (-\ell\pm 2(n\pm1)p)\al_+ + \ell'\al_-\right)
& for $n$ odd$\,,$\cr} \cr
\De_\pm^{(n)}(\ell,\ell') & = \half \left( p_\pm^{(n)}(\ell,\ell')
-Q \right)^2 -
\half Q^2 + \textstyle{{1\over16}}(1-2\ka)\,,\cr}
}
and
$\widetilde{E}_{m,m'}(p,p') = \{ \half - \De^{(n)}_+(m,m')\,,n\in\ZZ\}
$. Let $d(\De)=|n|$ for $\De = \half-\De_+^{(n)}(m,m')$ and
$\et(p^L) = {\rm sign}(i(p^L-Q^L))$. Under the plausible assumption that
the Felder resolution generalizes to the super-case
we have (this result was also announced in [\LZb])
\thm\PBETb
\def\LFF{ L(\De(m,m'))\otimes \cF^{sL}(p^L)\otimes \cF^{sG}}
\proclaim Theorem \PBETb.
\item{(i)} $
H^{(*)}_{rel}(\LFF) \neq0\quad {\rm iff}\quad \De(p^L)\in
\widetilde{E}_{m,m'}(p,p')\,,
$
\item{(ii)} $
{\rm dim\ } H^{(n)}_{rel}(\LFF) = \de_{n,\et(p^L)d(\De(p^L))}\,.
$
\par

\acknow
\bs

\noindent
{\it Note added:} While preparing this paper we received [\IO]
in which Theorem \PBETa\ is also presented.  The proof given there for
the NS sector is exactly as we discussed above.  However, since
$\widehat{d}^{\,2} \neq 0$ on the $L_0=0$ subspace of the R sector
($\widehat{d}=d_R$ in [\IO]), the discussion for this case seems
misleading.

\footatend\immediate\closeout\rfile\writestoppt
\baselineskip=14pt{\bigskip\noindent {\bf  7. References}}%
\bigskip{\frenchspacing%
\parindent=20pt\escapechar=` \input refs.tmp\vfill\eject}\nonfrenchspacing
\vfill\eject\end